\documentclass[prl,twocolumn,showpacs,superscriptaddress,nofootinbib,floatfix]{revtex4-2}
\usepackage[utf8]{inputenc}
\usepackage{amsmath,amssymb}
\usepackage{slashed}
\usepackage{subfigure}
\usepackage[colorlinks=true, 
linkcolor=blue,
breaklinks=true,
urlcolor=magenta,
citecolor=blue]{hyperref}
\usepackage[usenames,dvipsnames]{color}
\usepackage{appendix}
\usepackage{braket,bm}
\usepackage{multirow}
\usepackage{enumitem}
\usepackage{color}
\usepackage{cancel}
\usepackage{mathrsfs}

\newcommand{\phy}{\mathrm{phy}}
\newcommand{\lecs}{\mathrm{LECs}}
\newcommand{\asc}{a_{\mathrm{sc}}}

\usepackage{graphicx}
\usepackage{tikz}
\usepackage[normalem]{ulem}
\usepackage{booktabs}
\usepackage{array}
\usepackage{overpic}
\usepackage{float}
\usepackage{threeparttable}
\allowdisplaybreaks[4]

\graphicspath{{fig/}}

\usetikzlibrary{calc}
\usetikzlibrary{intersections}
\usetikzlibrary{trees}
\usetikzlibrary{decorations.pathmorphing}
\usetikzlibrary{decorations.markings}
\usetikzlibrary{arrows.meta}
\usetikzlibrary{patterns}

\newcommand{\hebtu}{\affiliation{Department of Physics and Hebei Key Laboratory of Photophysics Research and Application,\\
Hebei Normal University, Shijiazhuang 050024, China}}

\newcommand{\seu}{\affiliation{School of Physics, Southeast University, Nanjing 211189, China}}

\begin{document}
\title{Prominent enhancement of axion thermalization rate from axion-kaon interactions}

\author{Jin-Bao Wang}\email{230228550@seu.edu.cn}
\seu\hebtu

\author{Zhi-Hui Guo}\email{zhguo@hebtu.edu.cn}
\hebtu

\author{Hai-Qing Zhou}\email{zhouhq@seu.edu.cn}
\seu

\begin{abstract}
The axion thermalization rate is a crucial input to determine the hot dark matter bound of axions, resulting from the scattering processes in the thermal bath of early Universe. We demonstrate that the commonly employed axion thermalization rate by including the $a\pi \leftrightarrow \pi\pi$ channel alone is significantly underestimated for the temperature $T$ above 100~MeV. This is obtained through the systematical calculation of the axion-light flavor meson scattering amplitudes within the framework of the chiral unitarization approach, paying special attention to the $a K \leftrightarrow \pi K$ reaction. Hadron resonances appearing in $a K \leftrightarrow \pi K$ amplitudes significantly enlarge the cross sections, which turn out to be much bigger than that of $a\pi \leftrightarrow\pi\pi$. The axion thermalization rate is then substantially enhanced by the $a K \leftrightarrow \pi K$ channel for $T\gtrsim 100$~MeV. Especially at $T\simeq 130$~MeV, the contribution from the $a K \leftrightarrow\pi K$ reaction to the axion thermalization rate exceeds the $a\pi\leftrightarrow\pi\pi$ one. Obviously more stringent constraints on the axion parameters are obtained, when confronting the number of extra relativistic degrees of freedom $\Delta N_{\rm eff}$ from Planck$'$18. 
\end{abstract}
\maketitle

{\em Introduction.---} The Peccei-Quinn (PQ) mechanism elegantly resolves the strong $CP$ problem in quantum chromodynamics (QCD), by introducing the axion, a pseudo-Nambu-Goldstone boson (pNGB) arising from the spontaneously breaking of the global $\rm U(1)_{PQ}$ symmetry~\cite{Peccei:1977hh,Peccei:1977ur,Weinberg:1977ma,Wilczek:1977pj}. Interestingly, the intriguing axion particle kills two birds with one stone, because it also supplies a well motivated feasible dark matter candidate, compromising the whole or part of the dark matter compounds~\cite{Preskill:1982cy,Abbott:1982af,Dine:1982ah}. Consequently, tremendous efforts have been made to search for the imprints of axion, across many disciplines of physics, such as the astronomy, cosmology, the fifth-force detection, quantum spin sensors, haloscope cavity, etc~\cite{Kim:2008hd,Graham:2015ouw,Irastorza:2018dyq,Sikivie:2020zpn,Choi:2020rgn,DiLuzio:2020wdo,Cong:2024qly,Jiang:2024boi}. 

The characterized axion interaction is given by the $aG\widetilde{G}/f_a$ term, being $G$ and $\widetilde{G}$ the gluon field tensor and its dual, due to its featured role in explaining the absence of $CP$ violation in strong interaction. Due to the experimental constraints from beam dump facilities and rare meson decays~\cite{Donnelly:1978ty,Hall:1981bc,Wilczek:1977zn,Bardeen:1986yb,Kim:1986ax,Turner:1989vc}, the invisible axion scenario, viz. requiring the axion decay constant $f_a\gg v_{\rm EW}\simeq 246$~GeV, remains a feasible framework. This however indicates that axion would faintly interact with the Standard Model (SM) particles, which makes the quest for axion in terrestrial experiments quite challenging.    

Cosmological observations can play important roles in probing the axion dynamics. In spite of diverse production mechanisms of the axion dark matter, relativistic relics of axions can be abundantly generated through their scattering with SM particles in the thermal plasma in early Universe. The relativistic axions contribute to additional dark radiation, the amount of which can be conveniently quantified by $\Delta N_{\rm eff}$, viz. the effective number of extra relativistic degrees of freedom (d.o.f.). The surveys of cosmic microwave background (CMB) by the Planck Collaboration~\cite{Planck:2018vyg,Planck:2018jri} can probe such quantity to a high precision, which can be then exploited to set constraints on the axion decay constant $f_a$ or its mass $m_a\simeq 5.7~\mathrm{eV}(\frac{10^6~\mathrm{GeV}}{f_a})$~\cite{GrillidiCortona:2015jxo}. The axion decoupling temperature $T_D$, at which the axions dissociate from the primordial thermal bath and stream freely until today, plays the decisive role. When axions decouple early at high temperatures well above $T_c\simeq 160$~MeV, i.e., the critical temperature of QCD phase transition, the population of such axions will be much diluted and leave faint imprint, which may reach the sensitivity of the future CMB-S4 measurement~\cite{CMB-S4:2022ght}. On the other hand, the current CMB results from Planck$'$18~\cite{Planck:2018vyg,Planck:2018jri} are more relevant for the situation with $T_D<T_c$. Below $T_c$, the relevant thermalization channels of axions are the processes involving hadrons, such as $a\pi\leftrightarrow\pi\pi$, $aK\leftrightarrow\pi K$, $aN\leftrightarrow\pi N$, etc, as quarks and gluons are confined in hadrons in this circumstance.

As the lightest hadron, pion exhibits the largest particle number density in the thermal bath, whereas contributions from heavier hadrons to the axion thermalization rate suffer exponential suppression due to their dilute number densities. Thus, existing works~\cite{Chang:1993gm,Hannestad:2005df,Hannestad:2007dd,Melchiorri:2007cd,Giusarma:2014zza,DEramo:2021psx,DiLuzio:2021vjd,DiLuzio:2022gsc,Notari:2022ffe,Wang:2023xny} primarily assume the $a\pi\to\pi\pi$ reaction as the dominant thermalization channel below $T_c$. The validity of this assumption comes with a caveat: there is no large enhancement for the axion amplitude involving heavier hadrons compared to the $a\pi\to\pi\pi$ case. Such assumption has never been seriously examined in any previous study. 
In this work, we reassess the approximation of  exclusively considering $a\pi\to\pi\pi$ scattering by investigating contributions from heavier channels. We expect the most promising channel from the $aK\to\pi K$ scattering, due to these arguments: (1) kaon is the next-to-lightest hadron after pion; (2) the overall small isospin breaking (IB) factor appearing in the $a\pi\to\pi\pi$ amplitudes is absent in the $aK\to\pi K$ process~\cite{Notari:2022ffe}; (3) prominent hadron resonances in the $\pi K$ system, such as $K^*(892)$ and $K^*_0(700)$, can particularly enhance the $aK\to\pi K$ reaction rate. 
Regarding the channels involving $\eta$, such as $a\eta\to\pi\pi$, $a\pi\to\pi\eta$, $a\eta\to K\bar{K}$, $aK\to K \eta$ and so on, either the broad feature of the scalar resonance $f_0(500)$ or the heavier masses of $a_0(980)$ and $f_0(980)$ or the higher thresholds are expected to make these channels subdominant, compared to the $a K\to\pi K$ process. 
To properly account for the resonance effects in both $a\pi\to\pi\pi$ and $aK\to\pi K$ processes, the nonperturbative hadronic interactions will be systematically incorporated via the unitarization procedure, which takes the results from chiral perturbation theory ($\chi$PT) as inputs. 
Our study shows that the inclusion of the $a K\to \pi K$ scattering substantially enlarges the axion thermalization rate at temperatures above 100~MeV, in contrast with the situation by only including the $a\pi\to\pi\pi$ channel, which in turn leads to an obviously more stringent constraint to the axion parameters when confronting with the Planck$'$18 measurement on $\Delta N_{\rm eff}$.  

{\em Axion-light meson scattering.---} In order to consistently highlight the impact from the $aK\to \pi K$ scattering, compared to the $a\pi\to\pi\pi$ contribution, we will stick to the model-independent axion interaction operator $\alpha_s aG\widetilde{G}/(8\pi f_a)$, as widely used in previous works for the calculation of the $a\pi\to\pi\pi$ amplitude~\cite{Chang:1993gm,DiLuzio:2022gsc,Notari:2022ffe,Wang:2023xny}. $\rm SU(3)$ $\chi$PT~\cite{Gasser:1984gg} is needed to account for the axion-kaon interactions. Following the standard procedure to perform the axial transformation of the quark fields, $q\to e^{i\frac{a}{2f_a}\gamma_5 Q_a} q$, with the constraint on the trace $\langle Q_a\rangle=1$, one can obtain the LO $\chi$PT Lagrangian with axion
\begin{equation} 
\mathcal{L}_2=\frac{F_{\pi}^2}{4}\langle \partial_{\mu}U\partial^{\mu}U^{\dagger}+\chi(a)U^{\dagger}+U\chi^{\dagger}(a) \rangle
-\frac{\partial_{\mu}a}{2f_a}\sum_{i=1}^8C_iJ^{\mu}_{A,i}\,,\label{eq.L2}
\end{equation}
where $U=e^{\frac{i\phi}{F_{\pi}}}$ with $\phi=\sum_{i=1}^8\lambda_i\phi_i$ the octet matrix of light pseudoscalar mesons and $\lambda_i$ Gell-Mann matrices, and the pion decay constant $F_{\pi}=92.1$~MeV. The axion-dressed scalar source is $\chi(a)=2B_0e^{-i\frac{a}{2f_a}Q_a}M_q e^{-i\frac{a}{2f_a}Q_a}$, with $M_q={\rm diag}( m_u,m_d,m_s )$ the diagonal quark mass matrix. It is common to take $Q_a=M_q^{-1}/\langle M_q^{-1} \rangle$~\cite{Georgi:1986df}, so that the $a$-$\pi^3$ and $a$-$\eta_8$ mass mixing terms disappear at LO. The LO axial-vector current accompanied by coefficient $C_i=\langle Q_a\lambda_i \rangle$ is  $J^{\mu}_{A,i}=i\frac{F_{\pi}^2}{4}\langle \lambda_i\{\partial^{\mu}U,\,U^{\dagger}\} \rangle$.
Since the singlet component of axial currents is related to the QCD $\mathrm{U}_A(1)$ anomaly that cannot be systematically addressed in $\mathrm{SU}(3)$ $\chi$PT and plays a marginal role in the $\pi,K$ sector~\cite{Gasser:1984gg}, such effect is neglected. The coefficients of the nonisosinglet components of the axial currents are 
\begin{align}\label{eq.defc3c8}
C_3=\frac{z(1-r^2)}{2r+z(1+r)^2}\,,\quad C_8=\frac{z(1+r)^2-4r}{\sqrt{3}\left[2r+z(1+r)^2\right]}\,,
\end{align}
with $z=\frac{m_s}{\hat{m}}\,,r=\frac{m_u}{m_d}\,,\hat{m}=\frac{m_u+m_d}{2}$. 
Although the above two terms cause $a$-$\pi^3$ and $a$-$\eta_8$ kinetic mixing, their effects on axion interactions can be neglected as long as we only keep the terms up to $\mathcal{O}(\frac{1}{f_a})$. 
The lowest order nonderivative axion interaction relevant to the $aP_1\to P_2P_3$ process from Eq.~(\ref{eq.L2}) turns out to be $\frac{a}{f_a}i\frac{F_{\pi}^2}{2}C_S\langle U-U^{\dagger} \rangle$ with $C_S= \frac{B_0}{\langle M_q^{-1} \rangle}=\frac{m_{\pi}^2}{\frac{(1+r)^2}{r}+\frac{2}{z}}$.
One can then calculate all the relevant $aP_1\to P_2P_3$ amplitudes at LO, the calculation detail is given in the Appendix. 

Several recent works~\cite{DiLuzio:2021vjd,DiLuzio:2022gsc,Notari:2022ffe,Wang:2023xny} have independently confirmed that the realistic $a\pi\to\pi\pi$ amplitudes are clearly underestimated by the LO $\chi$PT results. The hadron resonances appearing in those reactions can obviously enlarge the axion thermalization rate. It is quite plausible that such enhancements will also happen in the $aK\to\pi K$ scattering. To consistently account for the hadron resonances in all the relevant $aP_1\to P_2 P_3$ reactions, we adopt the unitarization recipe in Refs.~\cite{Oller:1997ng,Oller:1998hw} that has been successfully applied to incorporating the nonperturbative interactions in the two-meson scattering processes by relying on the perturbative $\chi$PT as inputs. The unitarized two-meson partial-wave (PW) scattering amplitudes are given by 
\begin{align}
&T^{\rm uni}_{IJ}=T^{(2)}_{IJ}\ldotp\left[T^{(2)}_{IJ}-T^{(4)_\lecs}_{IJ}-T^{(2)}_{IJ}\ldotp\mathcal{G}\ldotp T^{(2)}_{IJ}\right]^{-1}\ldotp T^{(2)}_{IJ}\,,\label{eq.unit}
\end{align}
where $T^{(2)}_{IJ}$ denotes the $\mathcal{O}(p^2)$ $\chi$PT PW amplitude with isospin $I$ and angular momentum $J$, $T^{(4)_\lecs}_{IJ}$ stands for the polynomial parts with $\mathcal{O}(p^4)$ low energy constants (LECs)~\cite{Gasser:1984gg} and $\mathcal{G}$ is the two-point one-loop function evaluated in dimensional regularization with unknown subtraction constant~\cite{Oller:1998zr}, whose imaginary part is ${\rm Im \mathcal{G}}=q/(8\pi\sqrt{s})$ for $\sqrt{s}$ above the two-meson threshold, with $q$ the three-momentum in the center of mass (CM) frame. In the coupled-channel case, all the quantities in Eq.~\eqref{eq.unit} should be understood as matrices. It is then straightforward to verify the unitarity relation obeyed by the unitarized amplitudes~\eqref{eq.unit}, viz. ${\rm Im} T=T^\dagger\ldotp q/(8\pi\sqrt{s}) \ldotp T$.  
Regarding the $aP_1\to P_2 P_3$ reaction, its PW amplitude $M$ should fulfill the unitarity relation of ${\rm Im} \vec{M}=T^\dagger\ldotp q/(8\pi\sqrt{s}) \ldotp \vec{M}$. Following this principle, one can similarly construct the unitarized $aP_1\to P_2P_3$ PW amplitudes via 
\begin{align}
&\vec{M}^{\rm uni}_{IJ}=T^{(2)}_{IJ}\ldotp\left[T^{(2)}_{IJ}-T^{(4)_\lecs}_{IJ}-T^{(2)}_{IJ}\ldotp\mathcal{G}\ldotp T^{(2)}_{IJ}\right]^{-1}\ldotp\vec{M}^{(2)}_{IJ}\,,\label{eq.unim} 
\end{align}
where $\vec{M}^{(2)}_{IJ}$ stands for the $\mathcal{O}(p^2)$ $\chi$PT PW amplitude of the $aP_1\to P_2P_3$ process, with $I$ the isospin number of the $P_2 P_3$ system.
Note that axion interaction associated with $C_3$ in Eq.~\eqref{eq.defc3c8}, behaving roughly as $(m_d-m_u)/(m_d+m_u)$, breaks the isospin symmetry, and it serves as the largest source of IB in $aP_1\to P_2P_3$ reaction. Since the IB correction in the hadron system is generally expected at the level around 2\%, so the IB effect in the final-state strong interactions (FSIs) of the meson-meson system will be ignored. 
One can easily prove that the unitarized amplitudes in Eqs.~\eqref{eq.unit} and \eqref{eq.unim} satisfy the strict unitarity relation aforementioned. 

\begin{figure*}[t]
	\centering
	\includegraphics[width=\linewidth]{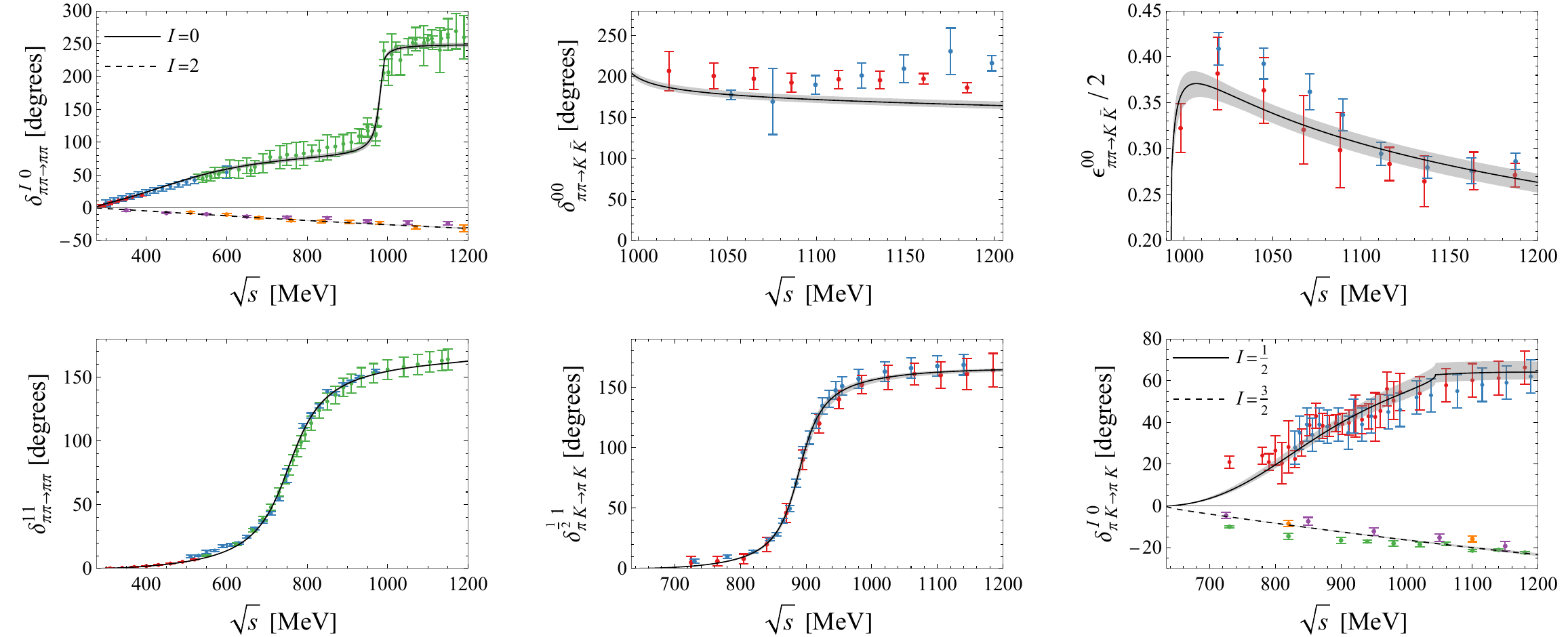}
	\caption{Results of the fits to scattering data. In the first row: the first panel shows the $\pi\pi\to\pi\pi$ phase shifts of $IJ=00$ (solid line) and $20$ (dashed line) with the data from \cite{NA482:2007xvj,Froggatt:1977hu,Guo:2011pa,Hoogland:1977kt,Losty:1973et}; the second panel shows $\delta_{\pi\pi\to K\bar{K}}^{00}$ with data from \cite{Cohen:1980cq,Etkin:1981sg}; the third panel shows the $\pi\pi\to K\bar{K}$ inelasticity of $IJ=00$ with data from \cite{Cohen:1980cq,Martin:1979gm}. In the second row: the first panel shows $\delta_{\pi\pi\to\pi\pi}^{11}$ with data from \cite{Garcia-Martin:2011iqs,Estabrooks:1974vu,Protopopescu:1973sh}; the second panel shows $\delta_{\pi K\to\pi K}^{1/2,1}$ with data from \cite{Mercer:1971kn,Estabrooks:1977xe}; the third panel shows $\delta_{\pi K\to\pi K}^{1/2,0}$ (solid line) with data from \cite{Guo:2011pa,Aston:1987ir}, along with $\delta_{\pi K\to\pi K}^{3/2,0}$ (dashed line) with data from \cite{Estabrooks:1977xe,Bakker:1970wg,Cho:1970fb}. The shaded areas surrounding each line stand for the uncertainties from the bootstrap method and this rule applies to all the figures below, see the text for details.}    
	\label{fig_scattering_data}
\end{figure*}

Given charge-conjugation symmetry, we only need to calculate four independent types of $a\pi\to\pi\pi$ and $aK\to\pi K$ processes, whose FSIs can be resummed by the above unitarization procedure. We incorporate the contributions from both $S$ and $P$ waves, and elaborate the interactions separately for the four kinds of reactions below: 
\begin{itemize}
	\item[(1)] $a\pi^0\to\pi^+\pi^-,\,\pi^0\pi^0$.  The isospin of $\pi\pi$ final states can only be $I=0,\,2$ due to $C$-parity conservation. Only $S$ wave needs to be considered. The $IJ=00$ channel contains the $f_0(500)$ and $f_0(980)$ resonances, while the $IJ=20$ is a nonresonant channel. The next heavier $K\bar{K}$ coupled state is also included for the proper description of FSIs with $IJ=00$. In $IJ=20$ sector, the elastic $\pi\pi$ interaction is sufficient for our study.
	\item[(2)] $a\pi^+\to\pi^+\pi^0$. Final state $\pi^+\pi^0$ can be decomposed into $IJ=11$ and $IJ=20$ channels. The former includes the prominent $\rho(770)$ resonance. $K^+\bar{K}^0$ coupled state is further introduced to describe FSIs in this channel. The $IJ=20$ channel shares the same FSIs with that in $a\pi^0$ case as we neglect the tiny IB effect in the two-meson FSIs. 
	\item[(3)] $a K^+\to\pi^+K^0,\,\pi^0K^+$. $\pi K$ in final states can be decomposed into the channels with $IJ=\frac{1}{2}0$, $\frac{1}{2}1$, $\frac{3}{2}0$, and $\frac{3}{2}1$. The former two are resonant channels which comprise $K^*_0(700)$/$\kappa$ and $K^*(892)$, respectively. We additionally introduce the coupled state $\eta K$ to describe the FSIs in the scattering processes with $IJ=\frac{1}{2}0$, $\frac{1}{2}1$. Excellent description for the $IJ=\frac{3}{2}0$ sector can be achieved by using $\pi K$ single-channel scattering. The interaction in the nonresonant channel with $IJ=\frac{3}{2}1$ is rather weak, and we will neglect its effect.
	\item[(4)] $aK^0\to\pi^0K^0,\,\pi^-K^+$. The situation of the FSIs in this case is the same as that of $aK^+$.
\end{itemize}

The parameters in the unitarized amplitudes include the subtraction constants $\asc$ in $\mathcal{G}$ and the $\mathcal{O}(p^4)$ LECs $\hat{L}_i$ in $T^{(4)_{\lecs}}_{IJ}$. Their values are determined through fitting the various meson-meson scattering data, including phase shifts and inelasticities. 
The uncertainty analyses are performed within the framework of bootstrap approach, by fitting the numerous pseudo-data sets generated via the random Gaussian sampling of the experimental scattering data. The large samples of refitted parameters resulting in this procedure are then used for later uncertainty studies throughout. 
The values of all the relevant parameters used in this work, together with their error bars, are presented in the Supplement. The fitted results are shown in Fig.~\ref{fig_scattering_data}.

\begin{figure*}[t]
    \centering
    \includegraphics[width=0.45\linewidth]{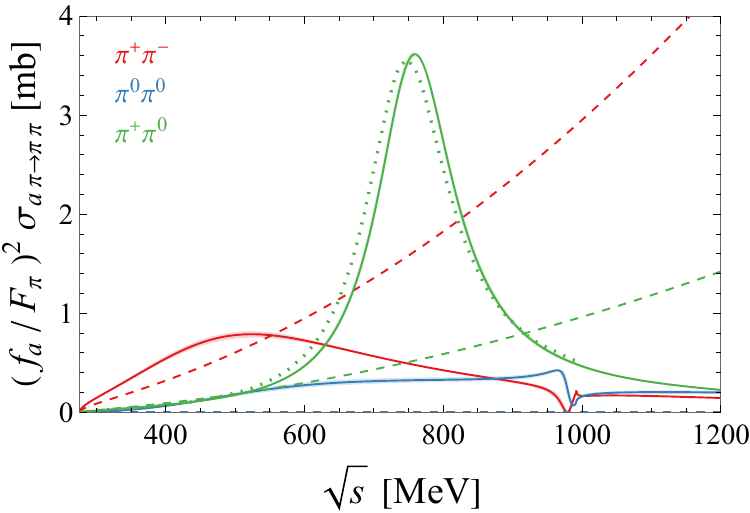}
    \includegraphics[width=0.45\linewidth]{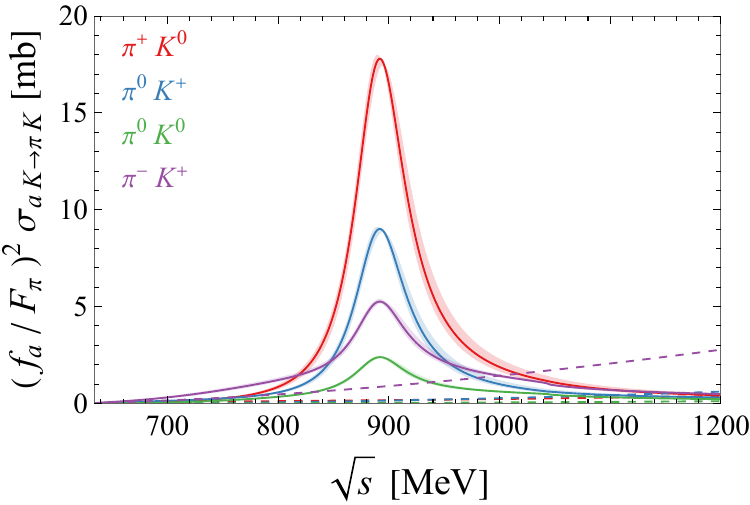}
    \caption{Cross sections for $a\pi\to\pi\pi$ processes (left) and $aK\to\pi K$ processes (right). Solid lines represent our central results by taking unitarized amplitudes, while dashed lines from LO amplitudes are only shown for comparison. In the left panel we also show the cross section of $a\pi^+\to\pi^+\pi^0$ scattering calculated by IAM amplitude~\cite{DiLuzio:2022gsc}, denoted by green dotted line.}
    \label{fig_cross_sections}
\end{figure*}

\begin{figure*}[t]
    \centering
    \includegraphics[width=0.47\linewidth]{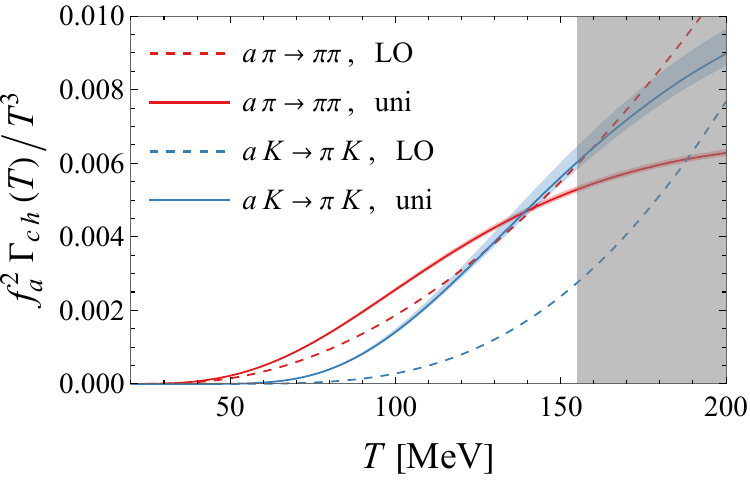}
    \includegraphics[width=0.45\linewidth]{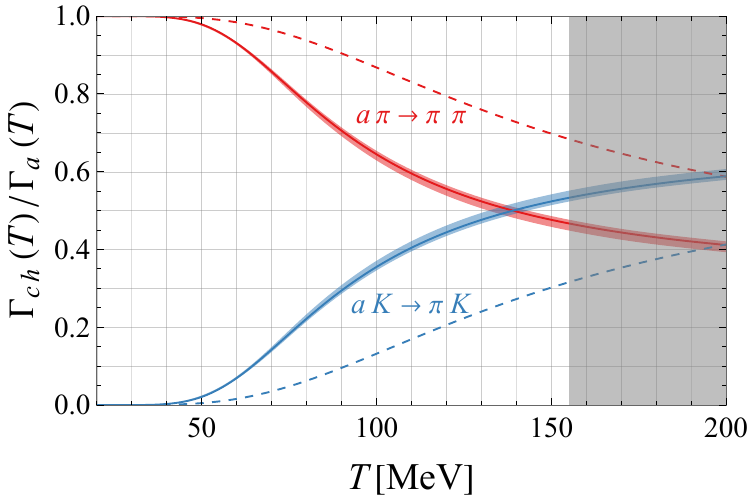}
    \caption{Magnitudes of the thermalization rates (left) and the relative contributions in percentage from $a\pi\to\pi\pi$ and $aK\to\pi K$ components (right). Solid lines represent our central results from unitarized amplitudes, while dashed lines from LO amplitudes are shown for comparison. For the regions above QCD crossover temperature, i.e., $T>155$~MeV, which are shaded in gray, the chiral amplitudes are no longer reliable, and they are simply shown to illustrate the growing trends of the kaon channels when increasing the temperatures.}
    \label{fig_thermalrate}
\end{figure*}

\begin{figure}[t]
    \centering
    \includegraphics[width=1.0\linewidth]{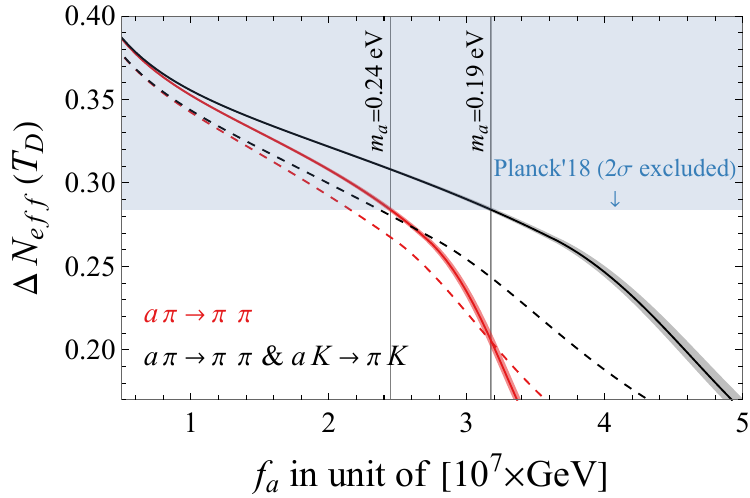}
    \caption{The constraint on the axion parameters $f_a$ and $m_a$ from $\Delta N_{\mathrm{eff}}$ of Planck$'$18~\cite{Planck:2018vyg}. The dashed and solid lines represent the results obtained by LO and unitarized amplitudes. The red curves only include $a\pi\to\pi\pi$ contribution, while the black curves include the additional $aK\to\pi K$ contribution. 
    }\label{fig_cosmologics}
\end{figure}

The unitarized $aP_1\to P_2P_3$ scattering amplitudes have many merits, such as   respecting chiral symmetry at low energies, satisfying exact unitarity relations and the constraints from the scattering data, and incorporating the dynamics of heavy d.o.f. by properly generating the relevant resonance states. These features ensure that the unitarized amplitudes correctly capture the key dynamics of $aP_1\to P_2P_3$ processes.
By taking those sophisticated unitarized PW amplitudes as inputs, we show the cross sections of $a\pi\to\pi\pi$ and $aK\to\pi K$ processes in CM frame in Fig.~\ref{fig_cross_sections}, evaluated through
\begin{equation}
\sigma_{i\to f}(s)=\frac{1}{32\pi s}\frac{|\vec{p}_f|}{|\vec{p}_i|}\int_{-1}^{+1}\mathrm{d}\cos\theta\left| M_{i\to f}(s,\cos\theta) \right|^2\,,
\end{equation}
where $\vec{p}_{i,f}$ stand for the three momenta of the initial and final states and $\theta$ is the scattering angle in the CM frame. 
The low-lying resonances appearing in these reactions, particularly $\rho(770)$ and $K^*(892)$, significantly enlarge the cross sections. The corresponding cross sections with LO amplitudes denoted by dashed lines in Fig.~\ref{fig_cross_sections} are only shown for the purpose of comparison, and they are invalid in the resonance energy regions. We take the reliable results from the unitarized amplitudes to proceed the following discussions. Strikingly, the sum of various $aK\to\pi K$ cross sections is obviously much larger than that of $a\pi\to\pi\pi$. In the left panel of Fig.~\ref{fig_cross_sections}, we also show the result of $a\pi^+\to\pi^+\pi^0$ calculated by the inverse amplitude method (IAM)~\cite{DiLuzio:2022gsc}, denoted by green dotted line. Our result is almost coincident with the IAM one, which serves as a crosscheck of the correctness of our approach.

{\em Axion thermalization rate.---}
The average rate of thermal production and annihilation of axions in the thermal bath, i.e., the axion thermalization rate, via the scattering channel denoted as $ch$, is given by
\begin{equation}
\begin{aligned}
\Gamma_{ch}(T)=\frac{1}{n_a^{eq}}\int\mathrm{d}\widetilde{\Gamma}_{ch}&\left|M_{ch}\right|^2n_B(E_1)n_B(E_2)
\\
\times&\left[1+n_B(E_3)\right]\left[1+n_B(E_4)\right]\,,
\end{aligned}\label{eq.rate}
\end{equation}
where $n_B(E)=1/\left(e^{E/T}-1\right)$, $n_a^{eq}=\zeta(3)T^3/\pi^2$ is the axion number density in thermal equilibrium, and the phase space integral is given by
\begin{equation}
\mathrm{d}\widetilde{\Gamma}_{ch}=\left[\prod_{i=1}^4\frac{\mathrm{d}^3p_i}{(2\pi)^32E_i}\right](2\pi)^4\delta^4(p_1+p_2-p_3-p_4)\,.\label{eq.phasespace}
\end{equation}
Due to rotational invariance, the integrand in Eq.~(\ref{eq.rate}) only depends on five independent kinematic variables. Accordingly, the 12-dimensional phase space integral in Eq.~(\ref{eq.phasespace}) can be reduced to a numerical integration over these five variables, with the detailed reduction procedure given in our previous work~\cite{Wang:2023xny}. 

The phase space integral in Eq.~(\ref{eq.phasespace}) spans the entire energy region. However, the unitarized amplitudes are expected to be reliable for $\sqrt{s}\leq1.2~$GeV, since we do not include scattering data at higher energies. This necessitates an evaluation of the reliability of axion thermalization rates computed using the unitarized amplitudes. We quantify this by computing the partial contribution $\Gamma^{\sqrt{s}\leq1.2~\mathrm{GeV}}_a(T)$, obtained by imposing a cutoff at $\sqrt{s}=1.2~$GeV. 
We verify that for temperatures below $150~$MeV, the $\sqrt{s}\leq1.2~$GeV region contributes over $96\%$ of the total thermalization rate. This dominant fraction substantiates the robustness of our results within this temperature regime. 

At temperatures below $T_c$, axions thermalize primarily through interactions with light-flavor hadrons in the SM thermal bath. In this temperature region, the axion-pion interaction is long assumed to be the only dominant channel in all the previous works~\cite{Chang:1993gm,Hannestad:2005df,Hannestad:2007dd,Melchiorri:2007cd,Giusarma:2014zza,DEramo:2021psx,DiLuzio:2021vjd,DiLuzio:2022gsc,Notari:2022ffe,Wang:2023xny}. The significantly large $a K\to \pi K$ cross sections as revealed in Fig.~\ref{fig_cross_sections} clearly challenge this assumption. The magnitudes and also relative contributions in percentage to the axion thermalization rate from $a\pi\to\pi\pi$ and $a K\to \pi K$ as a function of temperature are given in the left and right panels of Fig.~\ref{fig_thermalrate}, respectively. Notably, the $a K\to \pi K$ process starts to contribute more than 40\% of the total rate for $T\gtrsim 110$~MeV, and it exceeds the $a\pi\to\pi\pi$ contribution around $T\gtrsim 130$~MeV.  

The significant enhancement of axion thermalization rate due to the inclusion of the $a K\to \pi K$ process inevitably intensifies the hot dark matter (HDM) bound of axions. As a first application of this important result, we extract the axion decoupling temperatures $T_D$ through $\Gamma_a(T_D)=H(T_D)$ by taking different values of $f_a$, where $\Gamma_a(T)$ is the total axion thermalization rate and $H(T)$ is the Hubble rate. The contribution of thermal axions that decouple at $T_D$ to the effective number of extra relativistic d.o.f. is denoted by $\Delta N_{\mathrm{eff}}(T_D)$. The pertinent cosmological inputs used here are taken from~\cite{Saikawa:2018rcs}, which are also used in Refs.~\cite{DiLuzio:2021vjd,DiLuzio:2022gsc,Wang:2023xny}.
The bound of $\Delta N_{\mathrm{eff}}$ from Planck$'$18~\cite{Planck:2018vyg} on the axion decay constant $f_a$ is shown in Fig.~\ref{fig_cosmologics}. The enhanced axion thermalization rate resulting from both contributions of $a\pi\to\pi \pi$ and $aK\to\pi K$ leads to $f_a\geq 3.18_{-0.03}^{+0.04}\times 10^7$~GeV, compared to the bound $f_a\geq 2.45_{-0.02}^{+0.03}\times 10^7$~GeV obtained by only including $a\pi\to\pi\pi$. Clearly the additional contribution from the $a K \to \pi K$ channel tightens the lower limit of $f_a$ by approximately $30\%$. Such lower limit of $f_a$ can be converted to the upper bound of $m_a$ through $m_a^2=\frac{F_{\pi}^2}{f_a^2}C_S$ at LO accuracy, and the corresponding results are also indicted in Fig.~\ref{fig_cosmologics}. When using the next-to-leading order expression of $m_a$~\cite{Lu:2020rhp}, the bounds on $m_a$ will shift downward by about $2\%$. 

{\em Summary.---}  By properly implementing the nonperturbative two-meson interactions in the chiral unitarization procedure, we calculate the $a\pi\to\pi\pi$ and $aK\to\pi K$ scattering amplitudes that contain the prominent resonance contributions. The axion thermalization rate acquired by taking the unitarized $a K\to\pi K$ amplitudes is substantially enhanced for $T\gtrsim 100$~MeV, in contrast to the situation with only the $a\pi\to\pi\pi$ process. Our calculation invalidates the assumption that the $a\pi\to\pi\pi$ is the only dominant channel below $T_c$. Consequently, the HDM bound of axion parameter obtained by the Planck$'$18 constraint on $\Delta N_{\rm eff}$ is tightened by about $30\%$ after taking into account the $aK\to\pi K$ contribution, compared to only including $a\pi\to\pi\pi$ channel. 

\section*{Acknowledgements}
This work was supported in part by National Natural Science Foundation of China (NSFC) of China under Grants No.~12475078, No.~12150013, No.~11975090. Z.H.G is also partially supported by the Science Foundation of Hebei Normal University with Contract No.~L2023B09.

\bibliography{main}
\bibliographystyle{apsrev4-2}

\begin{widetext}
\newpage

\section{Appendix: Supplementary details about ``Prominent enhancement of axion thermalization rate from axion-kaon interactions''}

\setcounter{equation}{0}
\def\theequation{A.\arabic{equation}}

The axial-vector coupling terms in Eq.~(\ref{eq.L2}) give $a$-$\pi^3$ and $a$-$\eta_8$ kinetic mixing at LO, and there is also the $\pi^3$-$\eta_8$ mass mixing due to IB. Neglecting the $\mathcal{O}(\frac{1}{f_a^2})$ effects, the bilinear terms with neutral boson fields read 
\begin{equation}
\begin{aligned}
\mathcal{L}_{mix}=&\frac{1}{2}\partial_{\mu}a\partial^{\mu}a+\frac{1}{2}\partial_{\mu}\pi^3\partial^{\mu}\pi^3+\frac{1}{2}\partial_{\mu}\eta_8\partial^{\mu}\eta_8
+\varepsilon_F C_k^{a\pi} \partial_{\mu}a\partial^{\mu}\pi^3
+\varepsilon_F C_k^{a\eta} \partial_{\mu}a\partial^{\mu}\eta_8
\\
&-\frac{1}{2}m_{0\pi}^2\pi^3\pi^3-\frac{1}{2}m_{0\eta_8}^2\eta_8\eta_8+\frac{\Delta_I}{\sqrt{3}}\pi^3\eta_8\,,
\end{aligned}\label{eq.Lmix}
\end{equation}
with $\varepsilon_F=\frac{F_{\pi}}{f_a}$, $C_k^{a\pi(\eta)}=\frac{1}{2}C_{3(8)}$, 
$m_{0\pi}^2=B_0(m_u+m_d)\,, m_{0\eta_8}^2=\frac{1}{3}B_0(m_u+m_d+4m_s)$ and $\Delta_I=B_0(m_d-m_u)$. 
The $\pi^3$-$\eta_8$ mixing is a typical IB effect in strong interaction. We take $\Delta_I$ as an expansion parameter and keep terms up to $\mathcal{O}(\Delta_I)$. The mixing can be eliminated through the following field redefinitions:
\begin{subequations}
\begin{align}
&a=a_{\phy}-\varepsilon_F\left(C_k^{a\pi}-\Delta_IC_{k}^{a\eta}X_{\pi\eta}\right)\pi^0_{\phy}
-\varepsilon_F\left(C_k^{a\eta}+\Delta_IC_k^{a\pi}X_{\pi\eta}\right)\eta_{\phy}\,,\label{eq.red.a}
\\
&\pi^3=\pi^0_{\phy}+\Delta_I X_{\pi\eta}\eta_{\phy}\,,
\\
&\eta_8=\eta_{\phy}-\Delta_IX_{\pi\eta}\pi^0_{\phy}\,,\label{eq.red.eta}
\end{align}
\end{subequations}
with
\begin{equation}
X_{\pi\eta}=-\frac{1}{\sqrt{3}\left(m_{0\eta_8}^2-m_{0\pi}^2\right)}\,.
\end{equation}
Substituting Eqs.~(\ref{eq.red.a})--(\ref{eq.red.eta}) into Eq.~(\ref{eq.Lmix}), we can obtain the diagonal canonical quadratic form with masses unaffected by mixing up to $\mathcal{O}(\frac{1}{f_a})$ and $\mathcal{O}(\Delta_I)$. The masses at LO are:
\begin{equation}
m_{a}^2=\frac{F_{\pi}^2}{f_a^2}C_S\,,
\quad
m_{\pi}^2=m_{0\pi}^2\,,
\quad
m_{\eta}^2=m_{0\eta_8}^2\,,
\end{equation}
and $m_{K^{\pm}}^2=B_0(m_u+m_s)$, $m_{K^0}^2=B_0(m_d+m_s)$. Note that we neglect axion mass in the scattering amplitude as $m_a^2\sim \mathcal{O}(\frac{1}{f_a^2})$.

Substituting Eqs.~(\ref{eq.red.a})--(\ref{eq.red.eta}) into Eq.~\eqref{eq.L2}, one acquires the lowest order interaction Lagrangian for $aP_1\to P_2P_3$ processes 
\begin{align}
\mathcal{L}_{int}=&-\frac{1}{24F_{\pi}f_a}\partial_{\mu}a_{\phy}\sum_{i=3,8}C_i\langle \lambda_i(\partial^{\mu}\phi\phi^2-2\phi\partial^{\mu}\phi\phi+\phi^2\partial^{\mu}\phi) \rangle
+\frac{1}{6F_{\pi}f_a}C_Sa_{\phy}\langle\phi^3\rangle\,,
\end{align}
where
\begin{align}
\phi=\sum_{i=1}^8\lambda_i\phi_i=&\sum_{i=1,2,4\text{-}7}\lambda_i\phi_i
+\left(\lambda_3-\Delta_IX_{\pi\eta}\lambda_8\right)\pi^0_{\phy}
+\left(\lambda_8+\Delta_IX_{\pi\eta}\lambda_3\right)\eta_{\phy}\,.
\end{align}
Defining Mandelstam variables for $a(p_1)P_1(p_2)\to P_2(p_3)P_3(p_4)$ scattering: $s=(p_1+p_2)^2$, $t=(p_1-p_3)^2$, and $u=(p_1-p_4)^2$, the relevant LO amplitudes for $aP_1\to P_2P_3$ processes can be calculated and we give their expressions in the $Mathematica$ file~\cite{code}.

We decompose the meson-meson final states of the $aP_1\to P_2P_3$ scattering amplitudes into the isospin basis with the conventions: $|\pi^+\rangle=-|I=1,I_z=+1\rangle$, $|K^-\rangle=-|I=\frac{1}{2},I_z=-\frac{1}{2}\rangle$, and then construct the unitarized amplitudes according to Eqs.~(\ref{eq.unit}) and (\ref{eq.unim}) in terms of the PW amplitudes. The PW convention used in this work is
\begin{equation}
A_{i\to f}^J(s)=\frac{1}{2(\sqrt{2})^{N_i+N_f}}\int_{-1}^{+1}\mathrm{d}\cos\theta\,P_{J}(\cos\theta)A_{i\to f}(s,\cos\theta)\,,
\end{equation}
where $\theta$ is the scattering angle in CM frame and  $P_n(x)$ is Legendre polynomial. When the initial(final) state contains identical particles, $N_i(N_f)=1$, otherwise $N_i(N_f)=0$. The parameters in the unitarized amplitudes include the $\mathcal{O}(p^4)$ LECs $\hat{L}_i$ in $T^{(4)_{\lecs}}$ and the subtraction constants in the matrix $\mathcal{G}=\mathrm{diag}(G_n,\,G_m,\,\cdots)$, with 
\begin{equation}
\begin{aligned}
G_n(s)=G(\asc^n,s,m_{n_1},m_{n_2})=&
-\frac{1}{(4\pi)^2}\bigg[\asc^n-1+\log\frac{m_{n_2}^2}{\mu^2}+\frac{m_{n_1}^2-m_{n_2}^2+s}{2s}\log\frac{m_{n_1}^2}{m_{n_2}^2}
\\
&-\frac{\sqrt{\lambda(s,m_{n_1}^2,m_{n_2}^2)}}{s}\log\frac{m_{n_1}^2+m_{n_2}^2-s+\sqrt{\lambda(s,m_{n_1}^2,m_{n_2}^2)}}{2m_{n_1}m_{n_2}} \bigg]\,,
\end{aligned}
\end{equation}
being $\lambda(x,y,z)$ the standard K\"all\'en function. The coupled channels in our construction are elaborated in the main text. Accordingly, the subtraction constants include: $\asc^{\pi\pi,00}$ and $\asc^{K\bar{K},00}$ in $IJ=00$, $\asc^{\pi\pi,20}$ in $IJ=20$, $\asc^{\pi\pi,11}$ and $\asc^{K\bar{K},11}$ in $IJ=11$, $\asc^{\pi K,\frac{1}{2}0}$ and $\asc^{\eta K,\frac{1}{2}0}$ in $IJ=\frac{1}{2}0$, $\asc^{\pi K,\frac{1}{2}1}$ and $\asc^{\eta K,\frac{1}{2}1}$ in $IJ=\frac{1}{2}1$, $\asc^{\pi K,\frac{3}{2}0}$ in $IJ=\frac{3}{2}0$. We adopt the relations between subtraction constants to reduce the number of parameters in the fits: $\asc^{11}=\asc^{\pi\pi,11}=\asc^{K\bar{K},11}=\asc^{\pi\pi,20}=\asc^{\pi K,\frac{3}{2}0}$, $\asc^{\frac{1}{2}0}=\asc^{\pi K,\frac{1}{2}0}=\asc^{\eta K,\frac{1}{2}0}$, and $\asc^{\frac{1}{2}1}=\asc^{\pi K,\frac{1}{2}1}=\asc^{\eta K,\frac{1}{2}1}$. In the mathematica file~\cite{code}, we also give the expressions of the PW amplitudes of meson-meson scattering used in the unitarized amplitudes.

We use the unitarized meson-meson scattering PW amplitudes to fit the scattering data shown in Fig.~\ref{fig_scattering_data}. The PW $S$ matrix in our convention reads 
\begin{equation}
S_{IJ}=1+2i\sqrt{\frac{q}{8\pi\sqrt{s}}}\ldotp T^{uni}_{IJ}\ldotp \sqrt{\frac{q}{8\pi\sqrt{s}}}\,,
\end{equation}
from which we can extract the phase shifts and inelasticities by $(S_{IJ})_{kk}=\epsilon^{IJ}_{kk}e^{2i\delta^{IJ}_{kk}}$ ($k$ is not summed) and $(S_{IJ})_{kj}=i\epsilon^{IJ}_{kj}e^{i\delta^{IJ}_{kj}}$ for $k\neq j$. The independent 13 parameters and their fitted values to describe the scattering data are shown in Table~\ref{tab_parameters}.

\begin{table}
	\centering
	\caption{The fitted values of the parameters in unitarized amplitudes with $\chi^2/n_{\mathrm{d.o.f.}}=683.44/(279-13)$.}
	\label{tab_parameters}
	\renewcommand{\arraystretch}{1.2}
	\begin{tabular}{c|c|c|c}
		\hline\hline
		\multicolumn{2}{c|}{Subst. const.}&\multicolumn{2}{c}{Low energy constants}
		\\
		\hline
		$\asc^{\pi\pi,00}$&$-0.49^{+0.24}_{-0.23}$&$\hat{L}_1\times10^3$&$0.33^{+0.02}_{-0.02}$
		\\
		$\asc^{K\bar{K},00}$&$-1.51^{+0.20}_{-0.19}$&$\hat{L}_2\times10^3$&$0.97^{+0.05}_{-0.05}$
		\\
		$\asc^{11}$&$-1.38^{+0.33}_{-0.26}$&$\hat{L}_3\times10^3$&$-2.71^{+0.10}_{-0.11}$
		\\
		$\asc^{\frac{1}{2}0}$&$0.15^{+0.18}_{-0.21}$&$\hat{L}_4\times10^3$&$-0.77^{+0.09}_{-0.11}$
		\\
		$\asc^{\frac{1}{2}1}$&$1.53^{+0.76}_{-0.80}$&$\hat{L}_5\times10^3$&$3.51^{+1.39}_{-1.62}$
		\\
		&  &$\hat{L}_6\times10^3$&$-1.47^{+0.20}_{-0.24}$
		\\
		&  &$\hat{L}_7\times10^3$&$-0.77^{+0.24}_{-0.18}$
		\\
		&  &$\hat{L}_8\times10^3$&$4.05^{+0.37}_{-0.45}$
		\\
		\hline\hline
	\end{tabular}
\end{table}

In the numeric calculations, we do not distinguish the charge and neutral kaon masses. The meson masses are taken as follows: $m_{\pi}=138~\mathrm{MeV}$, $m_{K}=496~\mathrm{MeV}$, $m_{\eta}=548~\mathrm{MeV}$. The renormalization scale $\mu$ in $\mathcal{G}$ is taken to be $770$~MeV. We retain the $\pi^3$-$\eta_8$ mixing up to $\mathcal{O}(\Delta_I)$ in the LO axion-meson amplitudes, where $\Delta_I$ is estimated as: $\Delta_I\simeq(M_{K^0}^2)_{\mathrm{exp}}-(M_{K^{\pm}}^2)_{\mathrm{exp}}-\left[(M_{\pi^0}^2)_{\mathrm{exp}}-(M_{\pi^{\pm}}^2)_{\mathrm{exp}}\right]\simeq
498^2-494^2-(135^2-140^2)~\mathrm{MeV}^2=5343~\mathrm{MeV}^2$. We adopt the FLAG average of $N_f=2+1$ lattice results for the quark mass ratios~\cite{FlavourLatticeAveragingGroupFLAG:2024oxs} with $r=0.485$ and $z=27.42$ to estimate the values of $C_3$, $C_8$, and $C_S$, yielding $C_3=0.341$, $C_8=0.550$, and $C_S=0.216m_{\pi}^2$.

\end{widetext}

\end{document}